








\font\twelverm=cmr10 scaled 1200    \font\twelvei=cmmi10 scaled 1200
\font\twelvesy=cmsy10 scaled 1200   \font\twelveex=cmex10 scaled 1200
\font\twelvebf=cmbx10 scaled 1200   \font\twelvesl=cmsl10 scaled 1200
\font\twelvett=cmtt10 scaled 1200   \font\twelveit=cmti10 scaled 1200

\skewchar\twelvei='177   \skewchar\twelvesy='60


\def\twelvepoint{\normalbaselineskip=12.4pt
  \abovedisplayskip 12.4pt plus 3pt minus 9pt
  \belowdisplayskip 12.4pt plus 3pt minus 9pt
  \abovedisplayshortskip 0pt plus 3pt
  \belowdisplayshortskip 7.2pt plus 3pt minus 4pt
  \smallskipamount=3.6pt plus1.2pt minus1.2pt
  \medskipamount=7.2pt plus2.4pt minus2.4pt
  \bigskipamount=14.4pt plus4.8pt minus4.8pt
  \def\rm{\fam0\twelverm}          \def\it{\fam\itfam\twelveit}%
  \def\sl{\fam\slfam\twelvesl}     \def\bf{\fam\bffam\twelvebf}%
  \def\mit{\fam 1}                 \def\cal{\fam 2}%
  \def\tt{\twelvett}
  \textfont0=\twelverm   \scriptfont0=\tenrm   \scriptscriptfont0=\sevenrm
  \textfont1=\twelvei    \scriptfont1=\teni    \scriptscriptfont1=\seveni
  \textfont2=\twelvesy   \scriptfont2=\tensy   \scriptscriptfont2=\sevensy
  \textfont3=\twelveex   \scriptfont3=\twelveex  \scriptscriptfont3=\twelveex
  \textfont\itfam=\twelveit
  \textfont\slfam=\twelvesl
  \textfont\bffam=\twelvebf \scriptfont\bffam=\tenbf
  \scriptscriptfont\bffam=\sevenbf
  \normalbaselines\rm}



\def\beginlinemode{\endmode
  \begingroup\parskip=0pt \obeylines\def\\{\par}\def\endmode{\par\endgroup}}
\def\beginparmode{\endmode
  \begingroup \def\endmode{\par\endgroup}}
\let\endmode=\par
{\obeylines\gdef\
{}}
\def\singlespace{\baselineskip=\normalbaselineskip}

\def\oneandahalfspace{\baselineskip=\normalbaselineskip
  \multiply\baselineskip by 3 \divide\baselineskip by 2}
\def\doublespace{\baselineskip=\normalbaselineskip \multiply\baselineskip by 2}

\newcount\firstpageno
\firstpageno=2
\footline={\ifnum\pageno<\firstpageno{\hfil}
                                          \else{\hfil\twelverm\folio\hfil}\fi}
\let\rawfootnote=\footnote              
\def\footnote#1#2{{\rm\singlespace\parindent=0pt\rawfootnote{#1}{#2}}}
\def\raggedcenter{\leftskip=4em plus 12em \rightskip=\leftskip
  \parindent=0pt \parfillskip=0pt \spaceskip=.3333em \xspaceskip=.5em
  \pretolerance=9999 \tolerance=9999
  \hyphenpenalty=9999 \exhyphenpenalty=9999 }
\def\dateline{\rightline{\ifcase\month\or
  January\or February\or March\or April\or May\or June\or
  July\or August\or September\or October\or November\or December\fi
  \space\number\year}}
\def\received{\vskip 3pt plus 0.2fill
 \centerline{\sl (Received\space\ifcase\month\or
  January\or February\or March\or April\or May\or June\or
  July\or August\or September\or October\or November\or December\fi
  \qquad, \number\year)}}


\hsize=6.5truein
\vsize=8.9truein
\parskip=\medskipamount
\twelvepoint            
\doublespace            
\overfullrule=0pt       


\def\preprintno#1{
 \rightline{\rm #1}}    

\def\title                      
  {\null\vskip 3pt plus 0.2fill
   \beginlinemode \doublespace \raggedcenter \bf}

\def\author                     
  {\vskip 3pt plus 0.2fill \beginlinemode
   \singlespace \raggedcenter}

\def\affil                      
  {\vskip 3pt plus 0.1fill \beginlinemode
   \oneandahalfspace \raggedcenter \sl}

\def\abstract                   
  {\vskip 3pt plus 0.3fill \beginparmode
   \doublespace \narrower ABSTRACT: }

\def\endtitlepage               
  {\endpage                     
   \body}

\def\body                       
  {\beginparmode}               

\def\subhead#1{                 
  \vskip 0.25truein             
  {\raggedcenter #1 \par}
   \nobreak\vskip 0.25truein\nobreak}

\def\refto#1{$|{#1}$}           

\def\references                 
  {\subhead{References}        
   \beginparmode
   \frenchspacing \parindent=0pt \leftskip=1truecm
   \parskip=8pt plus 3pt \everypar{\hangindent=\parindent}}

\gdef\refis#1{\indent\hbox to 0pt{\hss#1.~}}    

\gdef\journal#1, #2, #3, 1#4#5#6{             
    {\sl #1~}{\bf #2}, #3, (1#4#5#6)}           

\def\refstylenp{                
  \gdef\refto##1{ [##1]}                                
  \gdef\refis##1{\indent\hbox to 0pt{\hss##1)~}}        
  \gdef\journal##1, ##2, ##3, ##4 {                     
     {\sl ##1~}{\bf ##2~}(##3) ##4 }}

\def\refstyleprnp{              
  \gdef\refto##1{ [##1]}                                
  \gdef\refis##1{\indent\hbox to 0pt{\hss##1)~}}        
  \gdef\journal##1, ##2, ##3, 1##4##5##6{               
    {\sl ##1~}{\bf ##2~}(1##4##5##6) ##3}}

\def\endreferences{\body}

\def\figurecaptions             
  { \beginparmode
   \subhead{Figure Captions}
}

\def\endpage                    
  {\vfill\eject}

\def\endpaper                   
  {\endmode\vfill\supereject}

\def\endit
  {\endpaper\end}


\def\ref#1{Ref. #1}                     
\def\Ref#1{Ref. #1}                     

\def\frac#1#2{{\textstyle{#1 \over #2}}}
\def\half{{\textstyle{ 1\over 2}}}

\def\sla{\raise.15ex\hbox{$/$}\kern-.57em}
\def\leaderfill{\leaders\hbox to 1em{\hss.\hss}\hfill}
\def\twiddle{\lower.9ex\rlap{$\kern-.1em\scriptstyle\sim$}}
\def\bigtwiddle{\lower1.ex\rlap{$\sim$}}
\def\gtwid{\mathrel{\raise.3ex\hbox{$>$\kern-.75em\lower1ex\hbox{$\sim$}}}}
\def\ltwid{\mathrel{\raise.3ex\hbox{$<$\kern-.75em\lower1ex\hbox{$\sim$}}}}
\def\square{\kern1pt\vbox{\hrule height 1.2pt\hbox{\vrule width 1.2pt\hskip 3pt
   \vbox{\vskip 6pt}\hskip 3pt\vrule width 0.6pt}\hrule height 0.6pt}\kern1pt}

\def\m@th{\mathsurround=0pt }
\def\leftrightarrowfill{$\m@th \mathord\leftarrow \mkern-6mu
 \cleaders\hbox{$\mkern-2mu \mathord- \mkern-2mu$}\hfill
 \mkern-6mu \mathord\rightarrow$}
\def\overleftrightarrow#1{\vbox{\ialign{##\crcr
     \leftrightarrowfill\crcr\noalign{\kern-1pt\nointerlineskip}
     $\hfil\displaystyle{#1}\hfil$\crcr}}}


\font\titlefont=cmr10 scaled\magstep3

\def\martinstyletitle                      
  {\null\vskip 3pt plus 0.2fill
   \beginlinemode \doublespace \raggedcenter \titlefont}

\font\twelvesc=cmcsc10 scaled 1200

\def\author                     
  {\vskip 3pt plus 0.2fill \beginlinemode
   \singlespace \raggedcenter\twelvesc}


\def\heading                            
  {\vskip 0.5truein plus 0.1truein      
   \beginparmode \def\\{\par} \parskip=0pt \singlespace \raggedcenter}

\def\subheading                         
  {\vskip 0.25truein plus 0.1truein     
   \beginlinemode \singlespace \parskip=0pt \def\\{\par}\raggedcenter}

\def\tag#1$${\eqno(#1)$$}

\def\align#1$${\eqalign{#1}$$}

\def\aligntag#1$${\gdef\tag##1\\{&(##1)\cr}\eqalignno{#1\\}$$
  \gdef\tag##1$${\eqno(##1)$$}}

\def\endaligntag{}

\def\overset #1\to#2{{\mathop{#2}\limits^{#1}}}
\def\underset#1\to#2{{\let\next=#1\mathpalette\undersetpalette#2}}
\def\undersetpalette#1#2{\vtop{\baselineskip0pt
\ialign{$\mathsurround=0pt #1\hfil##\hfil$\crcr#2\crcr\next\crcr}}}


\def\ref#1{Ref.~#1}                     
\def\Ref#1{Ref.~#1}                     
\def\[#1]{[\cite{#1}]}
\def\cite#1{{#1}}
\def\(#1){(\call{#1})}
\def\call#1{{#1}}
\def\taghead#1{}
\def\frac#1#2{{#1 \over #2}}
\def\half{{\frac 12}}

\def\12{{1\over2}}

\def\sla{\raise.15ex\hbox{$/$}\kern-.57em}
\def\leaderfill{\leaders\hbox to 1em{\hss.\hss}\hfill}
\def\twiddle{\lower.9ex\rlap{$\kern-.1em\scriptstyle\sim$}}
\def\bigtwiddle{\lower1.ex\rlap{$\sim$}}
\def\gtwid{\mathrel{\raise.3ex\hbox{$>$\kern-.75em\lower1ex\hbox{$\sim$}}}}
\def\ltwid{\mathrel{\raise.3ex\hbox{$<$\kern-.75em\lower1ex\hbox{$\sim$}}}}
\def\square{\kern1pt\vbox{\hrule height 1.2pt\hbox{\vrule width 1.2pt\hskip 3pt
   \vbox{\vskip 6pt}\hskip 3pt\vrule width 0.6pt}\hrule height 0.6pt}\kern1pt}
\def\tdot#1{\mathord{\mathop{#1}\limits^{\kern2pt\ldots}}}

\def\pmb#1{\setbox0=\hbox{#1}%
  \kern-.025em\copy0\kern-\wd0
  \kern  .05em\copy0\kern-\wd0
  \kern-.025em\raise.0433em\box0 }

\def\tifr{Tata Institute of Fundamental Research\\
Homi Bhabha Road, Bombay 400005, INDIA}

\def \r{\hat r}

\catcode`@=11
\newcount\tagnumber\tagnumber=0

\immediate\newwrite\eqnfile
\newif\if@qnfile\@qnfilefalse
\def\write@qn#1{}
\def\writenew@qn#1{}
\def\w@rnwrite#1{\write@qn{#1}\message{#1}}
\def\@rrwrite#1{\write@qn{#1}\errmessage{#1}}

\def\taghead#1{\gdef\t@ghead{#1}\global\tagnumber=0}
\def\t@ghead{}

\expandafter\def\csname @qnnum-3\endcsname
  {{\t@ghead\advance\tagnumber by -3\relax\number\tagnumber}}
\expandafter\def\csname @qnnum-2\endcsname
  {{\t@ghead\advance\tagnumber by -2\relax\number\tagnumber}}
\expandafter\def\csname @qnnum-1\endcsname
  {{\t@ghead\advance\tagnumber by -1\relax\number\tagnumber}}
\expandafter\def\csname @qnnum0\endcsname
  {\t@ghead\number\tagnumber}
\expandafter\def\csname @qnnum+1\endcsname
  {{\t@ghead\advance\tagnumber by 1\relax\number\tagnumber}}
\expandafter\def\csname @qnnum+2\endcsname
  {{\t@ghead\advance\tagnumber by 2\relax\number\tagnumber}}
\expandafter\def\csname @qnnum+3\endcsname
  {{\t@ghead\advance\tagnumber by 3\relax\number\tagnumber}}

\def\equationfile{%
  \@qnfiletrue\immediate\openout\eqnfile=\jobname.eqn%
  \def\write@qn##1{\if@qnfile\immediate\write\eqnfile{##1}\fi}
  \def\writenew@qn##1{\if@qnfile\immediate\write\eqnfile
    {\noexpand\tag{##1} = (\t@ghead\number\tagnumber)}\fi}
}

\def\callall#1{\xdef#1##1{#1{\noexpand\call{##1}}}}
\def\call#1{\each@rg\callr@nge{#1}}

\def\each@rg#1#2{{\let\thecsname=#1\expandafter\first@rg#2,\end,}}
\def\first@rg#1,{\thecsname{#1}\apply@rg}
\def\apply@rg#1,{\ifx\end#1\let\next=\relax%
\else,\thecsname{#1}\let\next=\apply@rg\fi\next}

\def\callr@nge#1{\calldor@nge#1-\end-}
\def\callr@ngeat#1\end-{#1}
\def\calldor@nge#1-#2-{\ifx\end#2\@qneatspace#1 %
  \else\calll@@p{#1}{#2}\callr@ngeat\fi}
\def\calll@@p#1#2{\ifnum#1>#2{\@rrwrite{Equation range #1-#2\space is bad.}
\errhelp{If you call a series of equations by the notation M-N, then M and
N must be integers, and N must be greater than or equal to M.}}\else%
{\count0=#1\count1=#2\advance\count1 by1\relax\expandafter\@qncall\the\count0,%
  \loop\advance\count0 by1\relax%
    \ifnum\count0<\count1,\expandafter\@qncall\the\count0,%
  \repeat}\fi}

\def\@qneatspace#1#2 {\@qncall#1#2,}
\def\@qncall#1,{\ifunc@lled{#1}{\def\next{#1}\ifx\next\empty\else
  \w@rnwrite{Equation number \noexpand\(>>#1<<) has not been defined yet.}
  >>#1<<\fi}\else\csname @qnnum#1\endcsname\fi}

\let\eqnono=\eqno
\def\eqno(#1){\tag#1}
\def\tag#1$${\eqnono(\displayt@g#1 )$$}

\def\aligntag#1\endaligntag
  $${\gdef\tag##1\\{&(##1 )\cr}\eqalignno{#1\\}$$
  \gdef\tag##1$${\eqnono(\displayt@g##1 )$$}}

\def\eqalignno#1{\displ@y \tabskip\centering
  \halign to\displaywidth{\hfil$\displaystyle{##}$\tabskip\z@skip
    &$\displaystyle{{}##}$\hfil\tabskip\centering
    &\llap{$\displayt@gpar##$}\tabskip\z@skip\crcr
    #1\crcr}}

\def\displayt@gpar(#1){(\displayt@g#1 )}

\def\displayt@g#1 {\rm\ifunc@lled{#1}\global\advance\tagnumber by1
        {\def\next{#1}\ifx\next\empty\else\expandafter
        \xdef\csname @qnnum#1\endcsname{\t@ghead\number\tagnumber}\fi}%
  \writenew@qn{#1}\t@ghead\number\tagnumber\else
        {\edef\next{\t@ghead\number\tagnumber}%
        \expandafter\ifx\csname @qnnum#1\endcsname\next\else
        \w@rnwrite{Equation \noexpand\tag{#1} is a duplicate number.}\fi}%
  \csname @qnnum#1\endcsname\fi}

\def\ifunc@lled#1{\expandafter\ifx\csname @qnnum#1\endcsname\relax}

\let\@qnend=\end\gdef\end{\if@qnfile
\immediate\write16{Equation numbers written on []\jobname.EQN.}\fi\@qnend}

\catcode`@=12

\catcode`@=11
\newcount\r@fcount \r@fcount=0
\newcount\r@fcurr
\immediate\newwrite\reffile
\newif\ifr@ffile\r@ffilefalse
\def\w@rnwrite#1{\ifr@ffile\immediate\write\reffile{#1}\fi\message{#1}}

\def\writer@f#1>>{}
\def\referencefile{
  \r@ffiletrue\immediate\openout\reffile=\jobname.ref%
  \def\writer@f##1>>{\ifr@ffile\immediate\write\reffile%
    {\noexpand\refis{##1} = \csname r@fnum##1\endcsname = %
     \expandafter\expandafter\expandafter\strip@t\expandafter%
     \meaning\csname r@ftext\csname r@fnum##1\endcsname\endcsname}\fi}%
  \def\strip@t##1>>{}}

\def\citeall#1{\xdef#1##1{#1{\noexpand\cite{##1}}}}
\def\cite#1{\each@rg\citer@nge{#1}}

\def\each@rg#1#2{{\let\thecsname=#1\expandafter\first@rg#2,\end,}}
\def\first@rg#1,{\thecsname{#1}\apply@rg}   
\def\apply@rg#1,{\ifx\end#1\let\next=\relax
\else,\thecsname{#1}\let\next=\apply@rg\fi\next}

\def\citer@nge#1{\citedor@nge#1-\end-}  
\def\citer@ngeat#1\end-{#1}
\def\citedor@nge#1-#2-{\ifx\end#2\r@featspace#1 
  \else\citel@@p{#1}{#2}\citer@ngeat\fi}    
\def\citel@@p#1#2{\ifnum#1>#2{\errmessage{Reference range #1-#2\space is bad.}%
    \errhelp{If you cite a series of references by the notation M-N, then M and
    N must be integers, and N must be greater than or equal to M.}}\else%
{\count0=#1\count1=#2\advance\count1 by1\relax\expandafter\r@fcite\the\count0,%
  \loop\advance\count0 by1\relax
    \ifnum\count0<\count1,\expandafter\r@fcite\the\count0,%
  \repeat}\fi}

\def\r@featspace#1#2 {\r@fcite#1#2,}    
\def\r@fcite#1,{\ifuncit@d{#1}
    \newr@f{#1}%
    \expandafter\gdef\csname r@ftext\number\r@fcount\endcsname%
                     {\message{Reference #1 to be supplied.}%
                      \writer@f#1>>#1 to be supplied.\par}%
 \fi%
 \csname r@fnum#1\endcsname}
\def\ifuncit@d#1{\expandafter\ifx\csname r@fnum#1\endcsname\relax}%
\def\newr@f#1{\global\advance\r@fcount by1%
    \expandafter\xdef\csname r@fnum#1\endcsname{\number\r@fcount}}

\let\r@fis=\refis           
\def\refis#1#2#3\par{\ifuncit@d{#1}
   \newr@f{#1}%
   \w@rnwrite{Reference #1=\number\r@fcount\space is not cited up to now.}\fi%
  \expandafter\gdef\csname r@ftext\csname r@fnum#1\endcsname\endcsname%
  {\writer@f#1>>#2#3\par}}

\def\ignoreuncited{
   \def\refis##1##2##3\par{\ifuncit@d{##1}%
    \else\expandafter\gdef\csname r@ftext\csname r@fnum##1\endcsname\endcsname%
     {\writer@f##1>>##2##3\par}\fi}}

\def\r@ferr{\endreferences\errmessage{I was expecting to see
\noexpand\endreferences before now;  I have inserted it here.}}
\let\r@ferences=\references
\def\references{\r@ferences\def\endmode{\r@ferr\par\endgroup}}

\let\endr@ferences=\endreferences
\def\endreferences{\r@fcurr=0
  {\loop\ifnum\r@fcurr<\r@fcount
    \advance\r@fcurr by 1\relax\expandafter\r@fis\expandafter{\number\r@fcurr}%
    \csname r@ftext\number\r@fcurr\endcsname%
  \repeat}\gdef\r@ferr{}\endr@ferences}


\let\r@fend=\endpaper\gdef\endpaper{\ifr@ffile
\immediate\write16{Cross References written on []\jobname.REF.}\fi\r@fend}

\catcode`@=12

\citeall\refto      
\citeall\ref        %
\citeall\Ref        %

\def\+{{\scriptscriptstyle +}}
\def\-{{\scriptscriptstyle -}}


\font\titlefont=cmr10 scaled \magstep3
\def\bigtitle{\null\vskip 3pt plus 0.2fill \beginlinemode \doublespace
\raggedcenter \titlefont}

\gdef\journal#1, #2, #3, 1#4#5#6{
{\sl #1~}{\bf #2}, #3 (1#4#5#6)}

\def\r{\rangle}

\def\bz{{\bar z}}
\def\bc{{\bar c}}
\def\bb{{\bar b}}

\def\bQ{{\bar Q}}
\def\PL{\Phi^{(L)}}
\def\PM{\Psi^{(M)}}
\def\dPL{\partial\PL}
\def\ddPL{\partial^2\PL}

\def\TG{T^{(G)}}

\def\TM{T^{(M)}}
\def\to{\rightarrow}
\def\half{{1\over2}}
\def\LL{L^{(L)}}
\def\LM{L^{(M)}}

\def\aL{\alpha^{(L)}}

\def\QL2{{Q_L\over2}}
\def\s2{\sqrt{2}}
\def\DMN{\Delta^{(M)(null)}}
\def\DLN{\Delta^{(L)(null)}}
\def\DM{\Delta^{(M)}}
\def\DL{\Delta^{(L)}}
\def\st{\sqrt{t}}

\def\PLNk{|\Psi^{(L)(null)}(k_L^\pm)\r}

\def\CK{{\cal K}}
\def\CC{{\cal C}}

\singlespace
\preprintno{TIFR/TH/91-50}
\preprintno{October 1991}
\doublespace
\bigtitle EXTRA STATES IN $c<1$ STRING THEORY$^*$
\bigskip
\author Sunil Mukhi
\affil\tifr
\abstract
A construction of elements of the BRS cohomology of ghost number $\pm 1$
in $c<1$ string theory is described, and their two-point function computed
on the sphere. The construction makes precise the relation between these
extra states and null vectors. The physical states of ghost number $+1$
are found to be exact forms with respect to a ``conjugate'' BRS operator.

\footnote{}{* Talk given at Carg\`ese Summer School, July 1991, based on
work done in collaboration with C. Imbimbo and S. Mahapatra}

\endtitlepage
\baselineskip = 16pt plus 1pt

\subhead{\bf 1. Introduction}

The no-ghost theorem\[thorn] for critical string theory states that
the BRS cohomology classes can be represented by
$$
c(z)\bc(\bz)~V(z,\bz)
\eqno(phys)
$$
where $V(z,\bz)$ is a dimension $(1,1)$ primary field of the $c=26$ matter
conformal field theory, $c(z)$ is the holomorphic spin $-1$ ghost and
$\bc(\bz)$ is its antiholomorphic counterpart.
There is an interesting exception to this theorem: the
identity field of the combined matter-ghost theory is certainly in the BRS
cohomology, but is not of the form given in Eq.\(0). Indeed, if the ghost
number of the states in Eq.\(phys) is chosen by convention to be $(0,0)$,
then that of the identity becomes $(-1,-1)$. In addition, one finds
a state of ghost number $(1,1)$ and a few more of mixed ghost number
$(1,0),(-1,0)$ and so on.

The chiral BRS operator for bosonic string theory in a given background is
$$
\eqalign{Q_B &= \oint dz~:c(z)\left(\TM(z) + \half\TG(z)\right):\cr
    &= \sum_{n = -\infty}^{\infty} c_n(\LM_{-n}
    - \half\sum_{m,n=-\infty}^{\infty} (m-n) :c_{-m} c_{-n} b_{m+n}:)\cr }
\eqno(brs)
$$
where $\TM(z), \TG(z)$ are the holomorphic stress-energy tensors of the
$c=26$ matter theory and the $c=-26$ ghost system respectively , and
$\LM_n$ are the modes of $\TM(z)$. In the closed string theory, we are
interested in the cohomology of $Q_B + \bQ_B$, the sum of the chiral and
antichiral BRS charges. More precisely, we must restrict to the cohomology
on the subspace annihilated by $b_0^- \equiv b_0 - \bb_0$ where $b(z)$ is
the chiral antighost field.  This is an example of a ``relative''
cohomology.

It can be shown that the cohomology for the closed string can be
reconstructed from a knowledge of that for the open string, in other
words, the cohomology of the chiral BRS operator alone. Even more, one can
restrict to the ``relative'' chiral cohomology, where we consider only the
subspace annihilated by $b_0$. The no-ghost theorem for the relative
chiral cohomology of the critical string says that the only physical
states are of the form $c_1|V\rangle$ where $|V\rangle$ is a chiral
primary state of dimension 1. The vacuum state $|0\rangle$, along with a
finite number of other states, provides an exception to the theorem. It is
important to note that all these states have zero $26$-momentum. Such
exceptional states are very few within the enormous classical phase space
of the critical string, and do not seem to be associated with significant
physical effects.

The situation is very different for non-critical string theories in
conformal backgrounds with $c\le 1$. In the present article I will
concentrate only on the case $c<1$, where the background matter theory is
a minimal model\[bpz]. In this case, the full Hilbert space is
obtained by taking the direct product of the matter CFT, the Liouville
theory (which is believed to describe the effect of quantized
two-dimensional gravity when the matter is non-critical), and the ghost
system. The BRS operator described above is generalized by adding the
Liouville stress-energy tensor to the matter one, wherever the latter
appears.

We parametrize the matter and Liouville central charges as
$$
\eqalign{c_M &= 13 - 6/t - 6t\cr
c_L &= 26 - c_M = 13 + 6/t + 6t\cr}
\eqno(cc)
$$
with $t=q/p~>0$. Here, $p$ and $q$ are two positive, coprime integers.
The Liouville stress-energy tensor is
$$
T^{(L)}(z) = -\half(\dPL\dPL + Q_L\ddPL)
\eqno(tl)
$$
where
$$
\eqalign{
Q_L &= \sqrt{{25 - c^{(M)}\over 3}}\cr
&=\s2(\sqrt{t} + 1/\sqrt{t})\cr}
\eqno(ql)
$$
The Virasoro generators following from this are
$$
\LL_n = \half\sum_{m=-\infty}^\infty :\aL_{n-m}\aL_m:
{}~+~~ {iQ\over2}(1+n)\aL_n
\eqno(ll)
$$
We follow the convention that Liouville vertex operators are defined as
$V^{(L)}_{k_L} =~:e^{k_L \PL}:$, with conformal dimension $\DL_{k_L} =
-\half k_L(k_L - Q_L)$ and $\aL_0$ eigenvalue $-i k_L$. Thus a Liouville
primary of given conformal dimension can have two possible values of
momentum, denoted $k_L^\pm$, where $k_L^+ > \QL2$, $k_L^- < \QL2$ and
$k_L^+ + k_L^- = Q_L$. We will denote the Fock space above a momentum
$k_L>\QL2$ as the ``$(+)$-Fock space'', and the other one as the
``$(-)$-Fock space''.

Then, a class of physical states in the (relative, chiral) cohomology is
again given by states like $c_1|V\rangle$, where this time $|V\rangle$ is
the direct product of a matter primary and a Liouville momentum state:
$$
|V\rangle = |\Psi\rangle_M \otimes |k_L\rangle_L
\eqno(state)
$$
Here, the Liouville momentum is adjusted such that the Liouville dimension
$\Delta_L$ and the matter primary dimension $\Delta_M$ satisfy $\Delta_L +
\Delta_M = 1$. Such states (two for every matter primary, because of the
two possible values of the Liouville momenta) are known as ``DDK states''
\[ddk].

Just like the critical string, the non-critical string also has
exceptional states which are not of the above kind, among which one
example is the vacuum state. However, it has been shown by Lian and
Zuckerman\[lz] that in this case there are {\it infinitely many}
exceptional physical states (we will call them ``LZ states'') in the
relative chiral cohomology, at {\it all} positive and negative values of
the ghost number. At the same time, the DDK states, analogous to the
dimension 1 primaries of the critical string, are finite in number for
minimal model backgrounds, since each minimal model has only finitely many
primary fields. Thus in these theories, the number of exceptional states
is a lot larger than that of the ``normal'' states, a situation quite
different from the critical string. One may expect these states to
play an important physical role in the analysis of non-critical string
theories.  (Various arguments have been put forward of late to support the
idea that there are actually infinitely many DDK states in minimal
backgrounds, because the decoupling of null vectors fails in the present
of gravity due to ``contact terms'' at boundaries of moduli space. It
remains true that the proportion of exceptional states of non-trivial ghost
number to the states of ghost number 0 is significantly greater than for
the critical string.)

The DDK states are often thought of as matter primaries ``dressed'' by a
Liouville momentum state to have total dimension 1. The principal
observation in \Ref{lz} is that given a primary field of the matter CFT,
exceptional physical states appear in the module whenever the Liouville
momentum is one which ``dresses'' a {\it null vector} over the matter
primary. In minimal models, each primary has an infinite chain of null
vectors over it. The ghost number of the physical state turns out to be
equal in magnitude to the distance of the associated null vector in this
chain from the original matter primary. The sign of the ghost number is
positive or negative if the Liouville momentum lies in the
$(+)$ or $(-)$ Fock space respectively. The dimension of the relative
chiral BRS cohomology is precisely 1 in every such case. Thus there is an
infinite set of physical states for each matter primary, one for each null
vector over it and for each sign of the ghost number.

In the rest of this article I will summarise an explicit method of
construction for a large class of LZ states, in arbitrary minimal models
coupled to gravity. The details, including proofs of a number of theorems,
can be found in Ref.\[imm].

\subhead{\bf 2. Null Vectors}

Degenerate fields in a $c<1$ CFT are those which have null vectors in the
Verma module above them. Their dimensions are given by the Kac formula:
$$
\DM_{r,s} = {(r^2 - 1)\over 4}{1\over t} + {(s^2 - 1)\over 4}t -
{(rs - 1)\over 2}
\eqno(kac)
$$
with $t = q/p$, ${1\le r \le q-1}$, ${1\le s\le p-1}$.

Each such field has infinitely many null vectors above it. We will refer
to the lowest of these null vectors in a module as ``primitive''.
The primitive $(r,s)$ null vector over $\Psi^{(M)}_{r,s}$ has dimension
$$
\eqalign{
\DMN_{r,s} &= \DM_{r,s} + rs\cr
&= {(r^2 - 1)\over 4}{1\over t} + {(s^2 - 1)\over 4}t +
{(rs + 1)\over 2}\cr }
\eqno(prnull)
$$

Now, an LZ state will occur whenever we consider a Liouville momentum
which ``dresses'' the above dimension:
$$
\eqalign{
\DL_{r,s} &= 1 - \DMN_{r,s}\cr
&={(1 - rs)\over 2} - {(r^2 - 1)\over 4}{1\over t} - {(s^2 - 1)\over 4}t\cr }
\eqno(dl)
$$
{}From the dimension formula above for Liouville vertex operators, it
follows that the two Liouville momenta are
$$
k_L^\pm = {1\over\s2} \left( {(1\pm r)\over\st} + (1\pm s)\st \right)
\eqno(kl)
$$
According to the Lian-Zuckerman theorem, a physical state of ghost number
$+1$ occurs in the Fock module above $k_L^+$, while a state of ghost
number $-1$ occurs above $k_L^-$.

There exists a rather neat formula, due to Benoit and Saint-Aubin\[bsa]
for the null vector in the Verma module of any CFT above a primary
satisfying the Kac formula\(kac) for the special case $r=1$, $s$
arbitrary:
$$
\eqalign{|\Psi^{(null)(Vir)}_{1,s}\r =
\sum_j~\sum_{p_1+p_2+\cdots+p_j=s,~p_i\ge 1}~&t^{s-j}
{{(s-1)!}^2 \over
\prod_{i=1}^{j-1}(p_1 + \cdots p_i)(r - p_1 - \cdots -p_i)}\cr
&L_{-p_1}L_{-p_2}\cdots L_{-p_j}|\Psi_{1,s}\r \cr }
\eqno(bsa)
$$
where $t$ parametrizes the central charge via the second equation in
Eq.\(cc). We will ultimately apply this to Liouville theory with $c>25$,
hence positive $t$.

One can ask how this null vector descends to the Fock space, for theories
with a Fock space description. It may in principle vanish identically when
re-expressed in oscillators, in which case it is in the kernel of the
projection map from the Verma module to Fock space. In this case, from
well-known arguments, there must be a state in the Fock space which is not
in the image of the projection. Alternatively, the projection to Fock
space may have no kernel in this module, in which case every state in the
Fock space lies in the image of the projection. This situation was
analyzed some years ago by Kato and Matsuda\[kato].  For our purposes we
need a stronger result than that of Ref.\[kato], namely:

\noindent{\bf Theorem 1:} For $t>0$,
$$
\eqalign{
&|\Psi^{(null)(Vir)}_{1,s}\r \to 0, ~~~ k_L < \QL2 \cr
&|\Psi^{(null)(Vir)}_{1,s}\r \to \prod_{k=1}^s (kt + 1)~
|\Psi^{(null)(Fock)}_{1,s}\r, ~~~ k_L > \QL2\cr }
$$
where the state $|\Psi^{(null)(Fock)}_{1,s}\r$ is a null state in the Fock
module which is non-vanishing for all values of $t$, and the arrow
indicates the projection map.

The proof of this theorem is given in Ref.\[imm]. It shows that in
general, for $t>0$, the projection to the $(-)$ Fock space has a
kernel, while the projection to the $(+)$ Fock space does not. But in
fact we learn something even for $t<0$. In this case, the result of
Theorem 1 holds with a possible interchange of $(+)$ and $(-)$ Fock
spaces, so that for the values $t=-1/k,~k=1,\cdots,s$, the projection to
{\it each} Fock space has a kernel. In other words, for these special
values of $t$, null vectors vanish when expressed in oscillators, in both
Fock spaces. This result will be crucial in the subsequent analysis. Note
that this in particular holds for $t=-1$ at every value of $s$, confirming
the well-known result that Virasoro null vectors in $c=1$ CFT vanish
identically in terms of oscillators.

\subhead{\bf 3. Construction of LZ States}

Although the LZ states occur in correspondence with matter null vectors,
they are not themselves null in any sense. They are to be found in the
module above a matter primary and a Liouville momentum which has the right
{\it dimension} to dress a matter null vector. This means that the
Liouville momentum $k_L$ corresponds to a conformal dimension $\DL$
which satisfies
$$
\DL + \DMN = 1
\eqno(null)
$$
where $\DMN$ is the dimension of some null vector above the chosen matter
primary.

Because of the well-known relation\[nullvectors] between null vectors for
two theories of central charge $c$ and $26-c$, we can re-state the above
result in a complementary way: whenever there is a Liouville null vector
(in the Fock module above a given momentum state) of dimension $\DLN$
satisfying
$$
\DLN + \DM =1
\eqno(nulltwo)
$$
for some matter primary of dimension $\DM$, the module contains an LZ
state.

To find these extra physical states, we start with a primitive $(1,s)$
Liouville null vector (in the Liouville Verma module) and the $(1,s)$
matter primary, combined into the state
$$ |X^{(0)}\r = \PLNk_L \otimes
|\PM_{1,s}\r_M \otimes c_1|0\r_G
\eqno(xo)
$$
The total dimension of this state is zero, by virtue of Eq.\(nulltwo)
and the fact that the ghost mode $c_1$ has dimension $-1$. It has the
same ghost number as DDK states, which we have chosen to call 0 by
convention.

We will construct physical states of ghost number $\pm 1$ starting from
the state defined above (this state is itself null, since its Liouville
sector is null.) Let us first define a conjugation operator $$
\CC:~~(+)~{\rm Fock~space}~\rightarrow~(-)~{\rm Fock~space}
\eqno(conj)
$$
which, acting on a state, simply replaces $k_L$ by $Q_L -k_L$. Clearly,
$\CC$ acts in either direction, and $\CC^2=1$. Using this operator we
define a ``conjugate'' BRS charge:
$$
Q_B^* \equiv \CC~Q_B~\CC
\eqno(qstar)
$$
which is nilpotent:
$$
(Q_B^*)^2=0
\eqno(qsquare)
$$
by virtue of the nilpotence of $Q_B$ and the fact that $\CC^2=1$.

Now, the state in Eq.\(xo) has the following property: because of
Theorem 1, $|X^{(0)}\r$ is in the kernel of the projection to the $(-)$
Fock space, so it vanishes identically when expressed in oscillators. On
the other hand, as long as $t\ne -1/k$ (in fact $t$ is positive for
Liouville theory), this state is not in the kernel of the projection to
the $(+)$ Fock space. Hence it descends to a non-vanishing Fock space
state, which is, however, both primary and secondary and hence null in the
usual sense. In particular this means that it is $Q_B$-exact, and hence of
course closed. Nevertheless, the operator $Q_B^*$ that we have just
defined has a non-trivial action on it. Indeed, define
$$
|X^{(1)+}\r \equiv Q_B^* |X^{(0)+}\r
\eqno(xone)
$$
Here, the $+$-superscript indicates that we are dealing with states in the
$(+)$ Fock space. The action of $Q_B^*$ is to first conjugate the state to
the $(-)$ Fock space, (where it becomes a non-primary, non-secondary state
with respect to the Virasoro algebra!), then act with $Q_B$, which
produces a nonzero result on such a state, and finally conjugate back to
the $(+)$ Fock space. The result is a state of ghost number $+1$ in the
$(+)$ Fock space, and we have:

\noindent {\bf Theorem 2:} The state $|X^{(1)+}\r$ defined in Eq.\(xone)
above is a LZ state of ghost number $+1$.

This is a remarkably simple result, and clarifies a conceptual point:
although LZ states are not themselves null, they are $Q_B^*$-variations of
null states. The fact that $Q_B^*$ has very different properties from $Q_B$
reflects a deep property of minimal matter coupled to gravity: the $(+)$
and $(-)$ Liouville Fock spaces are very different, a fact which has
played an important role in several
contexts\[kato]\[polch]\[seiberg]\[mms]\[imm].

Before discussing the proof of this theorem, we state the corresponding
result for the conjugate LZ state, which according to Ref.\[lz] should lie
in the $(-)$ Fock space and have ghost number $-1$. Let us define the
operator
$$
\CK \equiv \sum_{n=1}^\infty n c_{-n} c_n
\eqno(k)
$$
This operator is produced by anticommuting $Q_B$ and $Q_B^*$, as one can
easily check by explicit computation:
$$
\{Q_B, Q_B^*\}= (k_L^+ - k_L^-)^2 \CK
\eqno(qqstar)
$$
A property of $\CK$ that can be checked is that it has no kernel on
states of ghost number $-1$, hence its inverse exists on ghost number
$+1$, and is given by
$$
\CK^{-1}= \sum_{n=1}^\infty {1\over n} b_{-n} b_n
\eqno(kinv)
$$

Consider now the Fock space state
$$
|X^{(-1)+}\r=\CK^{-1} |X^{(1)+}\r
\eqno(xminusone)
$$
of ghost number $-1$. Since it is in the $(+)$ Fock space, where the
projection from the Verma module has no kernel, we can rewrite the left
hand side of this equation in terms of (Liouville and matter) Virasoro
secondaries acting on suitable primary states. Then, project this to the
$(-)$ Fock space. This procedure defines a state which we call
$|X^{(-1)-}\r$, of ghost number $-1$. This brings us to

\noindent {\bf Theorem 3:} The state $|X^{(-1)-}\r$ defined above is a LZ
state of ghost number $-1$.

Indeed, it is clear that $|X^{(-1)-}\r$ and $|X^{(1)+}\r$ are built on the
same Liouville primary, but the Liouville momenta are respectively $k_L^-$
and $k_L^+$, whose sum is precisely $Q_L$. Thus this pair of states can
have a non-vanishing inner product between them. The computation of this
inner product proves Theorems 2 and 3, according to which these states are
genuinely in the cohomology, for the following reason. The state
$|X^{(+1)+}\r$ is closed, because of its definition Eq.\(xone), the
anticommutation relation Eq.\(qqstar), and the fact that $|X^{(0)}\r$ is
annihilated by both $Q_B$ and $\CK$. This in turn implies that
$|X^{(-1)-}\r$ is closed, since $Q_B$ commutes with $\CK$ (an immediate
consequence of Eq.\(qqstar)). Now if both states are closed, and if their
inner product is nonvanishing, it follows that neither of them is exact,
which proves Theorem 2 and 3 above.  We find that in fact the inner
product is:
$$
\l X^{(-1)-}|c_0 |X^{(1)+}\r=s(s-1)!^2~\prod_{n=1}^{s-1} \left(
(nt)^2-1\right)
\eqno(inner)
$$
which is in fact nonvanishing for all $t$ other than $\pm 1, \pm{1\over 2},
\ldots, \pm{1\over s-1}$. Apart from the first pair of values, which
correspond to $c=1,25$, the remaining are ``unphysical'' minimal models.
Hence for all genuine $c<1$ minimal models, the inner product above is
nonvanishing, and our construction gives the LZ states associated to the
matter primaries of type $(1,s)$, at ghost number $\pm 1$. For $c=1$ it
does not work, a fact which may merit further investigation.

The detailed proof of Eq.\(inner), which is the principal result of this
work, is given in Ref.\[imm], along with various other expressions for the
physical states, and the inter-relations among them. Here I will just
sketch the proof. Starting from the Benoit-Saint-Aubin formula, Eq.\(bsa),
one can convince oneself that all the states that we have constructed
above, and their inner products, are polynomials in $t$. (This is not true
for null vectors of type $(r,s)$ for $r\ne 1$, where one finds polynomials
in $t$ and ${1\over t}$.) From the asymptotic behaviour of the formula,
one can evaluate the asymptotic behaviour, for large $t$, of the inner
product.  This suffices to fix the degree of the polynomial and the
leading coefficient. It remains only to determine the zeroes. Precisely
half of them are obtained from Theorem 1 above, since the special values
of $t$ at which the null vector vanishes on projecting to {\it both} Fock
spaces are clearly values for which the states we constructed above, and
hence their inner product, vanish. The remaining zeroes follow from the
fact that the transformation $t\rightarrow -t$ interchanges matter and
Liouville sectors, and one can argue that the inner products have a
definite parity under this transformation. This completes the derivation.

Although this derivation only works starting from the special null vectors
of type $(1,s)$, this appears to be only a technical limitation. One can
check in explicit examples that the same construction works also in
situations where $r\ne 1$, but a general proof for this is not available
at present. Another, more serious, limitation is the restriction to ghost
numbers $\pm1$.

\subhead{\bf 4. Conclusions}

More than a year after the discovery of infinitely many extra physical
states of every ghost number in the $c<1$ string, their physical
interpretation remains unclear. (Recently there has been very interesting
progress in the corresponding problem for the $c=1$
string\[witten]\[klebanov].) An understanding of the role played by these
states is likely to help clarify the situation regarding correlation
functions in $c<1$ strings, on which a lot of work has been done but a
clear understanding reconciling all the different approaches remains to be
achieved (see the lecture of V. Dotsenko in this volume, and references
therein.)

The present understanding of these correlation functions from the
continuum approach is based on analytic continuation and the eventual
insertion of a fractional and/or negative number of screening charges in
the Liouville sector. The fact that even vertex operators outside the
minimal table acquire non-vanishing correlators in this framework,
described by the heuristic expression that ``null vectors do not decouple
in the presence of gravity'', tends to agree with the qualitative feature
that there are infinitely many scaling fields in the matrix model and
topological approaches. On the other hand, we now have infinitely many LZ
physical states, which one might also be tempted to identify with this
infinity of scaling fields. It needs to be understood whether these states
are in some sense an alternate representation of the null vectors which
``do not decouple'', or should be interpreted in some different way.
(Some aspects of the relation between matrix model and continuum CFT
fields are discussed in Ref.\[mooress].)

The present work is an attempt not at addressing this question directly,
but rather at formulating and analysing the explicit form of LZ states,
which are much more non-trivial to write down than DDK states. An explicit
algorithm (quite distinct from the ``brute-force'' method that one can
always use) was obtained to construct the extra LZ states.  Explicit
examples are worked out in Ref.\[imm]. The algebraic structure related to
the conjugate $BRS$ operator $Q_B^*$ remains somewhat mysterious, and
perhaps once this is clarified then the physical interpretation of LZ
states for $c<1$ will become more evident.

\subhead{\bf 5. Acknowledgements} I wish to thank the organizers of the
Carg\`ese School for making it possible for me to participate and for their
kind invitation to give this talk. I am grateful to C. Imbimbo, S.
Mahapatra, S. Mukherji and A. Sen for their collaboration and for many
useful discussions.

\references

\refis{thorn} C. Thorn, Nucl. Phys. {\bf B286} (1987) 61.

\refis{bpz} A.A. Belavin, A.M. Polyakov and A.B. Zamolodchikov, Nucl.
Phys. {\bf B241} (1984) 333.

\refis{ddk} F. David, Mod. Phys. Lett. {\bf A3} (1988) 1651;\hfill\break
J. Distler and H. Kawai, Nucl. Phys. {\bf B321} (1989) 509.

\refis{lz} B.H. Lian and G.J. Zuckerman, Phys. Lett. {\bf B254} (1991) 417.

\refis{imm} C. Imbimbo, S. Mahapatra and S. Mukhi, Genova/Tata Institute
Preprint GEF-TH 8/91, TIFR/TH/91-27, May 1991.

\refis{bsa} L. Benoit and Y. Saint-Aubin, Phys. Lett. {\bf B215} (1988) 517.

\refis{kato} M. Kato and S. Matsuda, in ``Conformal Field Theory and
Solvable Lattice Models'', Advanced Studies in Pure Mathematics {\bf 16},
Ed. M. Jimbo, T. Miwa and A. Tsuchiya (Kinokuniya, 1988).

\refis{nullvectors} B.L. Feigin and D.B. Fuchs, Funct. Anal. and
Appl. {\bf 16} (1982) 114;

\refis{polch} J. Polchinski, Texas preprint UTTG-39-90.

\refis{seiberg} N. Seiberg, Rutgers Preprint RU-90-29 (1990).

\refis{mms} S. Mukherji, S. Mukhi and A. Sen, Phys. Lett. {\bf B266} (1991)
337.

\refis{witten} E. Witten, IAS Princeton preprint IASSNS-HEP-91/51, August
1991.

\refis{klebanov} I.R. Klebanov and A.M. Polyakov, Princeton University
preprint PUPT-1281,\hfil\break September 1991.

\refis{mooress} G. Moore, N. Seiberg and M. Staudacher, Rutgers-Yale
preprint RU-91-11, YCTP-P11-91.

\endreferences
\endit